\documentclass[prb,twocolumn,amsmath,amssymb,showpacs]{revtex4}
\newcommand{\beq}{\begin{equation}}
\newcommand{\eeq}{\end{equation}}

\usepackage{graphicx}
\usepackage{dcolumn}
\usepackage{bm}

\begin{document}   
\title{Large isotope effect on $T_c$ in cuprates despite of a small 
electron-phonon
coupling}

\author{Roland Zeyher}
\affiliation{Max-Planck-Institut f\"ur Festk\"orperforschung,
             Heisenbergstrasse 1, D-70569 Stuttgart, Germany}
\author{Andr\'{e}s Greco}
\affiliation{Departamento de F\'{\i}sica, Facultad de Ciencas Exactas
e Ingenier\'{\i}a and IFIR(UNR-CONICET), 2000 Rosario, Argentina}

\date{\today}

\begin{abstract}
We calculate the isotope coefficients $\alpha$ and $\alpha^\ast$
for the superconducting critical temperature $T_c$ and the pseudogap
temperature $T^\ast$ in a mean-field treatment of the $t$-$J$ model 
including phonons. The pseudogap phase is identified
with the $d$-charge-density wave ($d$-CDW) phase in this model.  
Using the small electron-phonon coupling constant $\lambda_d \sim 0.02$ 
obtained previously in LDA calculations in YBa$_2$Cu$_3$O$_7$,
$\alpha^{\ast}$
is negative but negligible small  whereas $\alpha$ increases 
from about 0.03 at optimal doping to values around 1 at small dopings 
in agreement with the general trend observed in many cuprates. 
Using a simple phase fluctuation
model where the $d$-CDW has only short-range correlations  it is
shown that the large increase of $\alpha$ at low dopings is rather 
universal and does not depend on the existence of sharp peaks in the density of
states in the pseudogap state or on specific values of the phonon cutoff.
It rather is caused by the large depletion of spectral weight at
low frequencies by the $d$-CDW and thus should also occur in other
realizations of the pseudogap.
\end{abstract}


\maketitle

\section{Introduction}
An open and rather controversially discussed topic in high-$T_c$
superconductivity is the role played by phonons for the 
low-energy electronic properties and, in particular, the high transition 
temperatures $T_c$. A direct way to show the involvement of the lattice
in electronic properties is the study of the isotope effect on 
$T_c$. \cite{Franck} Experimentally, the corresponding isotope
coefficient $\alpha$ is very
small in high-$T_c$ cuprates near optimal doping but increases strongly
in the underdoped region attaining values being comparable or even larger
than those in conventional phonon-mediated superconductors. It is often   
argued that these large observed isotope effects in the underdoped 
region give direct evidence for a large
electron-phonon coupling 
in the cuprates. \cite{Franck,Keller,Bornemann} 
It even has been suggested that it is so large that 
Eliashberg theory breaks down and that non-adiabatic and polaronic features 
play an important role in these systems. 
\cite{Alexandrov,Buss,Alexandrov1,Jarrell}
Other approaches have suggested that large isotope effects may occur
in the presence of a pseudogap. \cite{Williams,Dahm} 

Below we show that the essential features of the isotope experiments on T$_c$
can be explained within a mean-field approximation of the $t$-$J$ model
\cite{Morse,Hsu,Cappelluti,Benfatto}
using very small values for the electron-phonon coupling.
The pure $t$-$J$ model exhibits in mean-field
approximation a competition of $d$-wave superconductivity and a $d$-CDW
with transition temperatures $T_c$ and $T^\ast$, respectively. The
observed pseudogap can be identified with the $d$-CDW phase in this 
model. \cite{Cappelluti,Chakravarty} Whereas many experiments support the
idea of two competing phases \cite{Tacon,Lee,Wen,Kondo,Yu,Mook,Liu}
the nature of the additional
phase remains unclear and many proposals besides of the $d$-CDW have been
considered. \cite{Norman} Experiments suggest that its order parameter
has $d$-wave symmetry like that of the superconducting phase. This favors
an unconventional charge- or spin density wave state with internal $d$-wave 
symmetry rather than a conventional one with (anisotropic)
$s$-wave symmetry. The $t$-$J$ model
yields at large $N$ ($N$ is the number of spin components) such a
$d$-CDW but its relevance for the physical case $N=2$, for instance
in form of a phase without long-range but strong $d$-wave short-range order, 
remains unclear. \cite{Leung,Macridin,Oles}

In section II we introduce our model, its phase diagram and numerical
results for the competing superconducting and CDW order parameters,
and the corresponding transition
temperatures $T_c$ and $T^\ast$. We then add phonons assuming always 
that the electron-phonon interaction is very small so that they can be 
treated in the weak-coupling approximation. Explicit formulas for the
isotope coefficients $\alpha$ and $\alpha^\ast$ related to $T_c$ and $T^\ast$ 
will be given. In section III we present numerical results
for the doping dependence of $\alpha$ and $\alpha^\ast$.
In section IV we extend our treatment by including off-diagonal fluctuations
in the $d$-CDW state using the method of Refs. \cite{Lee1,Bartosch,Kuchinskii}.
In this way the pseudogap
phase is treated more realistically because the long-range order is 
removed and self-energy effects are included. Our conclusions are 
found in section V.

\section{Theoretical framework}

We consider the $t$-$J$ model \cite{Ogata} with the Hamiltonian $H$,
\begin{eqnarray} 
H = -t\sum_{\langle i,j \rangle,\sigma} c^\dagger_{i\sigma} c_{j\sigma}
-t'\sum_{\langle i,j \rangle',\sigma} c^\dagger_{i\sigma} c_{j\sigma}
\nonumber \\
+ \frac{J}{2} \sum_{\langle i,j \rangle} {\bf S}_i {\bf S}_j 
-\frac{1}{4}(J-2V_C)\sum_{\langle i,j \rangle} n_i n_j.
\label{H}
\end{eqnarray}
$c^\dagger_{i\sigma}, c_{i\sigma}$ are creation and 
annihilation operators, respectively, for electrons at site $i$ and
spin projection $\sigma$ subject to the condition that double occupancies 
of sites are excluded. 
The sums include nearest neighbor $\langle i,j \rangle$ and next nearest
neighbor $\langle i,j \rangle'$ sites on a 2D square lattice, the 
corresponding hopping elements are $t$ and $t'$, respectively. 
${\bf S}_i$ and $n_i$ are spin and site occupation operators, 
$J$ the Heisenberg coupling constant, and $V_C$ a Coulomb interaction
between nearest neighbors. 

One way to obtain a mean-field approximation
for $H$ is to introduce $N$ spin components in Eq. (\ref{H}), scale
the coupling constants as $t \rightarrow 2t/N$, $t' \rightarrow 2t'/N$,
$J \rightarrow 2J/N$ etc., and to consider the large $N$ 
limit. \cite{comment}
As a result $t$ and $t'$ are renormalized yielding the
quasi-particle dispersion $\epsilon({\bf k})$. At the same time the fermionic 
operators can be treated as usual creation and annihilation operators.  
Explicitly, one obtains 
$\epsilon({\bf k}) = -2(\delta t + r J)(\cos(k_x)+\cos(k_y))
-4t'\delta \cos(k_x)\cos(k_y)  -\mu$,
with $r = 1/N_c \sum_{\bf q} \cos({q_x})f(\epsilon({\bf q}))$.
$f$ is the Fermi function, $\delta$ the doping away from half-filling, and
$\mu$ a renormalized chemical potential. Here and in the following
we use the lattice constant $a$ of the square lattice as length unit.
As previously discussed \cite{Cappelluti,Greco} the 
relevant order parameters for our mean-field treatment of the $t$-$J$ model
is a CDW order parameter
\begin{equation}
\Phi({\bf k})=-\frac{i}{N_c}\sum_{\bf q}J({\bf k}-{\bf q})
\langle c^\dagger_{{\bf q}\uparrow} c_{{\bf q}+
{\bf Q}{\uparrow}} \rangle,
\label{J} 
\end{equation}
with $J({\bf k}) = 2J(\cos k_x +\cos k_y)$,
and a superconducting (SC) order parameter
\begin{equation} 
\Delta({\bf k}) = \frac{1}{N_c}\sum_{\bf q} 
(J({\bf k}-{\bf q})-V_C({\bf k}-{\bf q}))
\langle c_{{\bf q}\uparrow} c_{{-\bf q}\downarrow}\rangle. 
\label{Delta}
\end{equation}
\begin{figure}[b] 
\vspace*{-0.5cm}
\centerline{\includegraphics[angle=-90,width=9.0cm]{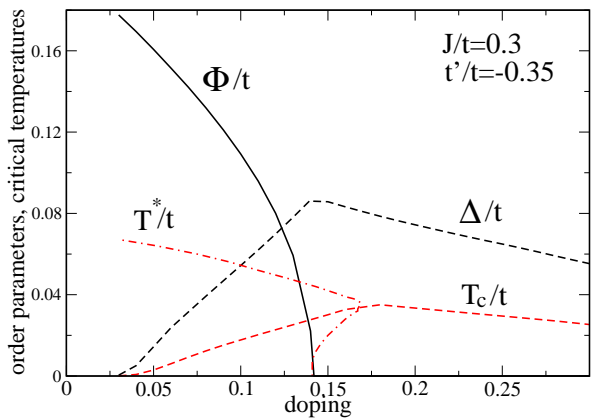}}
\vspace{-0.3cm}
\caption{\label{fig:1}
(Color online)
Zero temperature order parameters $\Phi$ and $\Delta$ and the 
critical temperatures $T^\star$ and $T_c$ as a function of doping.
}
\end{figure}
$N_c$ is the number of
primitive cells, $\langle...\rangle$ denotes an expectation value,
${\bf Q} =(\pi,\pi)$ is the wave vector of the $d$-CDW, and
$V_C({\bf k}) = 2V_C(\cos k_x +\cos k_y)$. The Coulomb interaction
$V_C({\bf k})$ between nearest neighbors has been introduced to prevent 
an instability of the $d$-CDW with respect to phase separation in some
region of phase space. \cite{Cappelluti}
From the self-consistency 
condition for the self-energy one obtains
coupled equations for $\Phi$, $\Delta$, the chemical potential and 
a renormalization contribution $r$  
to the band dispersion due to $J$. \cite{Cappelluti,Greco}
Their most stable 
solutions have $d$-wave symmetries in the interesting doping region,
i.e., $\Phi({\bf k})=\Phi \gamma({\bf k})$ and $\Delta({\bf k}) =
\Delta \gamma({\bf k})$ with $\gamma({\bf k})=(\cos k_x-\cos k_y)/2$.

A similar mean-field approximation as above is obtained by
using a slave-boson representation for $H$ in Eq.(\ref{H}),
enforcing the constraint on the average, using usual 
mean-field decouplings for the third and fourth term in $H$, and
dropping the antiferromagnetic order parameter. The above expression
for $\epsilon({\bf k})$ as well as Eq.(\ref{Delta}) are in this way exactly
reproduced, Eq.(\ref{J}) with $J/2+V_C$ instead of $J$.  

Fig. \ref{fig:1} shows the doping dependence of $\Phi$ and 
$\Delta$ at zero temperature,
calculated fully self-consistently for $t'/t=-0.35$, $J/t=0.3$, and 
$V_C/J=0.2$. In the overdoped 
region $\delta \geq \delta_c \sim 0.14$ $\Phi$ is zero. In the underdoped 
region $\delta \leq \delta_c $ $\Phi$ is non-zero and coexists 
with $\Delta$. $\Delta$  shows 
a maximum near $\delta_c$ and decays approximately
linearly in $\delta$ towards lower and higher dopings. 
The two order parameters compete with each other which causes the strong
decay of $\Delta$ with decreasing doping in the underdoped region. 
Also shown in Fig. \ref{fig:1} is $T^\star$ (dash-dotted line) and 
$T_c$ (dashed line) where $\Phi$ and $\Delta$, respectively, vanish.  
$T^\star$ shows near $\delta_c$ a reentrant behavior which has a simple
explanation: The piece of the $T^\star$ line above the $T_c$ curve 
is unaffected by superconductivity. Discarding superconductivity 
the $T^\star$ line would continue to the right decreasing slowly and
reaching the $x$ axis only at around $\delta \sim 0.25$. 
Taking superconductivity 
into account $T^\star$ and also $\Phi$ are suppressed by
the presence of a finite $\Delta$, i.e., below the $T_c$ curve.
Since $\Delta$ increases rapidly with
decreasing temperature $T^\star$ even bends back due to the strong
repulsion and reaches the critical doping $\delta_c$ at zero
temperature where $\Phi$ becomes nonzero. The reentrant behavior
thus reflects the strong competition of the CDW and SC order 
parameters. 
The occurrence of a large coexistence region of $\Delta$ and $\Phi$, 
which extends down to
$\delta = 0$,  is plausible because the 
Fermi surface consists in the $d$-CDW state of arcs around the nodal 
direction \cite{Greco}
which are unstable against the formation of a BCS gap $\Delta$. 

In order to discuss the isotope effect we consider 
a phonon-induced electronic density-density coupling between
nearest neighbors and on the same atom. Approximating 
its frequency dependence by a rectangular form, as is often done
in approximate solutions of the Eliashberg equation,\cite{Carbotte}
this effective
electron-electron interaction has in the $d$-wave channel the form,  
\begin{equation}
v({\bf q},i\omega_n) = -2V n_d({\bf q},i\omega_n) n_d({\bf -q},-i\omega_n),
\label{v}
\end{equation}
\begin{equation}
n_d({\bf q},i\omega_n) = \frac{1}{N_c}\sum_{{\bf k},\sigma=1,2}
\gamma({\bf k})\Theta_n c^\dagger_{{\bf k}+{\bf q}{\sigma}} 
c_{{\bf k}{\sigma}}.
\end{equation}
Similarly, we have in the isotropic $s$-wave channel,
\begin{equation}
w({\bf q},i\omega_n) = -\frac{W}{2} n({\bf q},i\omega_n) n({\bf -q},-i\omega_n),
\label{w}
\end{equation}
\begin{equation}
n({\bf q},i\omega_n) = \frac{1}{N_c}\sum_{{\bf k},\sigma=1,2}
\Theta_n c^\dagger_{{\bf k}+{\bf q}{\sigma}} 
c_{{\bf k}{\sigma}}. 
\end{equation}
$V$ and $W$ are electron-phonon (EP) coupling constants in the $d$
and $s$-wave channels, respectively,
$\Theta_n$ the cutoff function $\Theta(\omega_D-|\omega_n|)$, 
$\omega_n$ the bosonic Matsubara frequency $\omega_n= 2n\pi T$, and
$\omega_D$ the phonon cutoff frequency.
$n_d({\bf q},i\omega_n)$ and  $n({\bf q},i\omega_n)$ are
electronic density operators with $d$ and $s$-wave symmetry, respectively.  
Effects due to a small EP interaction can be taken into
account in the curves of Fig. 1 by adding the electronic self-energy  due
to $v({\bf q},i\omega_n)$ and $w({\bf q},i\omega_n)$ in the form of a 
Fock diagram. 
The resulting self-consistent equations lead to an equation for
the renormalization function $Z({\bf k},i\omega_n)$ due to $W$.
At $T=T_c$ this equation can be solved directly yielding
$Z({\bf k},i\omega_n) \equiv Z = 1 + \lambda_s$ where $\lambda_s$
is the product of $W$ and the electronic density at the Fermi energy
and $T=T_c$, i.e., it refers in general to the $d$-CDW state. A second 
equation is obtained
which determines the SC order parameter $\Delta({\bf k},i\omega_n)$,
\begin{eqnarray}
\Delta({\bf k},i\omega_n) = 
-4\tilde{J}\gamma({\bf k})\frac{T}{N_c}\sum_{{\bf k}'n'}
\gamma({\bf k}')g_{12}({\bf k}',i\omega_{n'}) - \nonumber\\
-4V\gamma({\bf k}) \Theta_n 
\frac{T}{N_c}\sum_{{\bf k}'n'}\gamma({\bf k}')
\Theta_{n'}g_{12}({\bf k}',i\omega_{n'}).
\label{gap}
\end{eqnarray}
$\tilde{J}$ is equal to $J-V_{C}$.
$g_{12}$ is the 12-element of the 4$\times$4
matrix Green's function $g$. Its inverse $g^{-1}({\bf k},i\omega_n)$
is given by,
\begin{equation}
\left( 
\begin{array}{c c c c}  
i\omega_nZ-\epsilon({\bf k}) & -\Delta({\bf k},i\omega_n)  & -i\Phi({\bf k},
i\omega_n)                      &  0                 \\
-\Delta({\bf k},i\omega_n)    &i\omega_nZ+\epsilon({\bf k})
&   0               &i\Phi({\bf \bar{k}},i\omega_n)      \\
i\Phi({\bf k},i\omega_n)      &   0
&i\omega_nZ-\epsilon({\bf\bar{k}})& -\Delta({\bf \bar{k}},i\omega_n)  
 \\
          0         &-i\Phi({\bf \bar{k}},i\omega_n)
&-\Delta({\bf\bar{k}},i\omega_n)&i\omega_nZ+\epsilon({\bf \bar{k}})
\end{array} \right)
\label{matrix}
\end{equation}
with the abbreviation
$\bf {\bar{k}}= {\bf k-Q}$. 
$\Phi({\bf k},i \omega_n  )$ is the
$d$-CDW order parameter renormalized by the phonons and given by,
\begin{eqnarray}
i \Phi({\bf k},i\omega_n) = -4J\gamma({\bf k}) \frac{T}{N_c} \sum_{{\bf k'}n'}
\gamma({\bf k'})g_{13}({\bf k'},i\omega_{n'}) \nonumber \\
+2V\gamma({\bf k}) \Theta_n \frac{T}{N_c}\sum_{{\bf k'}n'}
\Theta_{n'}\gamma({\bf k'})g_{13}({\bf k'},i\omega_{n'}),
\label{Phi}
\end{eqnarray}
where $g_{13}$ is the 13-element of the Green's function matrix $g$.

  For the calculation of $T_c$ it is sufficient to linearize the
right-hand side of Eq. (\ref{gap}) with respect to  $\Delta({\bf k},i\omega_n)$.
Furthermore, we may neglect the phonon renormalization for $\Phi$ in this
case: Below we will be interested only in a small EP constant $V$
yielding also only a small renormalization. Moreover, as will be shown below,
this small renormalization is practically independent of the ionic
mass $M$ and thus may be neglected in calculating the isotope effect on
$T_c$. We therefore have solved Eq. (\ref{Phi}) using only the first term 
on the right-hand side and use the solution in 
Eq. (\ref{matrix}) to obtain $g_{12}$. 
Eq. (\ref{gap}) represents an integral equation with two separable kernels
which can easily be solved. Writing Eq.(\ref{gap}) as a condition for $T_c$
we find,
\begin{equation}
(1+F_{11})(1+F_{22}) -F_{12}^2=0,
\label{Tc}
\end{equation}
with 
\begin{equation}
F_{11} = -2{\tilde J}\int_0^\infty d \omega 
\frac{N_d(Z\omega)}{Z\omega} tanh(\frac{\omega}{2T_c}),
\label{F11}
\end{equation}
\begin{eqnarray}
F_{12} = -2\sqrt{V{\tilde J}}\int_0^\infty d \omega \frac{N_d(Z\omega)}
{Z\omega}\frac{2}{\pi}
\Im[\psi(\frac{1}{2}+{\frac{i\omega}{2\pi T_c}})  \nonumber \\
  -\psi({\frac{\omega_D}{2\pi T_c}}+1 +{\frac{i\omega}{2\pi T_c}})],
\label{F12}
\end{eqnarray}
and $F_{22}=\sqrt{V/{\tilde J}}F_{12}$. $\psi$ is the digamma function and $\Im$
denotes the imaginary part.
$N_d({\omega})$ is given by
\begin{equation}
N_d(\omega) = \frac{2\omega}{\pi N_c} \sum_{\bf k} \gamma^2({\bf k})
\Im [G({\bf k},\omega-i\eta)],
\label{N}
\end{equation}
where $\eta$ is a positive infinitesimal quantity,
\begin{equation}
 G({\bf k},z) = \frac{z^2-\epsilon^2({\bf  \bar{k}})
-\Phi^2({\bf k})}{(z^2-\lambda_1^2({\bf k}))(z^2-\lambda_2^2({\bf k}))},
\label{Pole}
\end{equation}
and
\begin{equation}
\lambda_{1,2} = \frac{\epsilon({\bf k})+\epsilon({\bf \bar{k}})}{2} 
\pm \frac{1}{2}\sqrt{(\epsilon({\bf k})-\epsilon({\bf \bar{k}}))^2
+4\Phi^2({\bf k})}.
\label{Pole1}
\end{equation}  
\begin{figure}[t] 
\vspace*{-0.3cm}
\includegraphics[angle=270,width=9.cm]{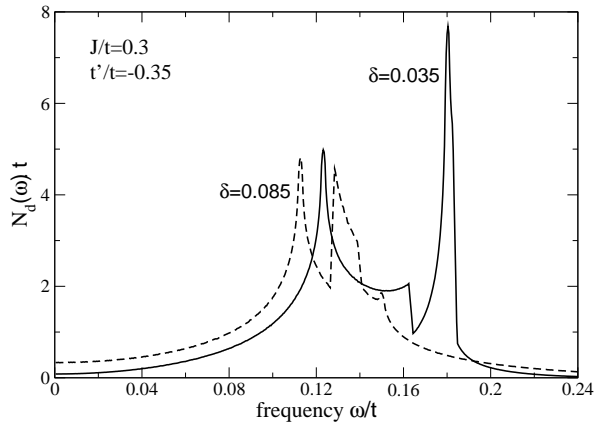}
\caption{\label{fig:2}
Weighted density $N_d(\omega)$ of electronic states for two dopings $\delta$. 
}
\vspace{-0.3cm}
\end{figure}
$N_d(\omega)$ represents a weighted density of electronic states at $T_c$
and is shown in Fig. \ref{fig:2} for two different dopings. It consists of a 
sharp peak near the energy $\Phi$ due to excitations across the $d$-CDW gap
and a structure at lower energies related to the van Hove singularity.
It is convenient to characterize $V$ by a dimensionless EP coupling constant,
\begin{equation}
\lambda_d = V N_d(0).
\label{lambdad}
\end{equation}

According to the above equations phonons affect $T_c$ in a two-fold
way, namely, via $V=\lambda_d/N_d(0)$ and via $Z=1+\lambda_s$. 
$\lambda_d$ and $\lambda_s$ characterize
the phonon-induced pairing interaction of $d$-wave and $s$-wave
symmetry, respectively. Putting $\lambda_s$ to zero $V$ increases $T_c$.
To see this we rewrite Eq. (\ref{Tc}) in the form $1+\hat{F}_{11}=0$.
$\hat{F}_{11}$ is given by Eq. (\ref{F11}) if one makes there the change
${\tilde J} \rightarrow {\tilde J}/(1-F_{12}^2/(1+F_{22}))$. 
This means that $V$ increases
${\tilde J}$ and thus increases $T_c$. On the other hand, if we put $V=0$
Eq. (\ref{Tc}) reduces to $1+\hat{F}_{11}=0$ with $\hat{F}_{11}$
given by Eq. (\ref{F11}) modified by ${\tilde J} \rightarrow {\tilde J}/Z$ and
$T_c \rightarrow T_c Z$. Each of these two changes diminishes $T_c$.
Thus phonons may lower or may increase $T_c$ depending which of 
the above two effects is larger. Numerical calculations indicate that
generically the second effect dominates and that $T_c$ decreases if
one couples to phonons. \cite{Macridin1} Most important for us is, however, 
the following observation. Our aim is not to determine the change in $T_c$
when the electron-phonon coupling is turned on but
when the ionic mass $M$ is changed. It is well known that $\lambda_s$
is independent of the ionic mass $M$, thus there will be no change
in $Z$ by isotope substitutions and $\alpha$ will always be positive. 
For small EP couplings we may even
put $Z=1$ and  keep only the linear term in $V$ 
in the calculation of the isotope coefficient  $\alpha =
-d \ln T_c /d \ln M$.    
From Eq. (\ref{Tc}) one finds for $\alpha$ in this limit,
\begin{equation}
\alpha = \frac{\omega_D}{T_c} F_{12} \Bigl(\frac{\partial F_{12}}{\partial
\omega_D} \Bigr) \Bigl(\frac{\partial F_{11}}{\partial T_c}\Bigr)^{-1},
\label{alpha}
\end{equation}
where the derivatives in Eq. (\ref{alpha}) are to be taken at the
$T_c$ without phonons and we also assumed $\omega_D
\sim M^{-0.5}$. 

\section{Results for the isotope coefficients}

In deriving the above formulas we assumed that the phonon-induced 
interaction $V$ affects only $\Delta$ but not $\Phi$ and thus also 
not $T^\ast$. To check this approximation we have calculated
the isotope coefficient $\alpha^\star$ related to $T^\star$ 
and defined by $\alpha^{\ast} = -d \ln T^{\ast} /d \ln M$.    
The calculation of $\alpha^{\ast}$ is very similar to that of $\alpha$.
\vspace{-0.2cm}
\begin{table}[h]
\caption{Isotope coefficient $\alpha^{\ast}$ for different dopings $\delta$
}
\vspace{0.3cm}
\begin{tabular}{c|c|c|c|c|c|c|c}
 $\delta$ & 0.028 & 0.048 & 0.064 & 0.090 & 0.115 & 0.139 & 0.164\\  
\hline 
$100\cdot\alpha^{\ast}$  & $-0.20$ & $-0.24$ & $-0.27$ & $-0.34$ & $-0.42$
 & $-0.53$ & $-0.66$
\end{tabular}
\end{table}
Numerical values for $\alpha^{\ast}$ as a function of doping throughout the
underdoped regime are given in TABLE I for $V/t=0.04$ and $\omega_D/t=0.1$. 
All values for
$\alpha^{\ast}$ are negative, i.e., $\alpha^{\ast}$ shows an inverse
isotope effect. However, this isotope effect is two orders of
magnitude smaller than the usual BCS value of 1/2 and thus tiny. Furthermore,
the absolute value of $\alpha^{\ast}$ decreases with decreasing doping
quite in contrast to $\alpha$ as will be shown below.
The negligible isotope effect on $T^\ast$ which we found
is in agreement with the experiment \cite{Williams} though there
exist also data which have been interpreted in terms of a large
isotope coefficient $\alpha^{\ast}$. \cite{Rubio}
Strictly speaking in our calculation of $\alpha$ 
the renormalized $d$-CDW order parameter at $T = T_c$ enters the 
density of states function $N_d(\omega)$, Eq. (\ref{F11}). Since we have
shown that $T^{\ast}$ is independent of the ionic mass $M$ 
we may conclude that the $d$-CDW order parameter at $T = T_c$ 
has also only a negligible isotope effect 
justifying the above procedure to calculate $\alpha$.

\begin{figure}[h] 
\vspace*{-0.2cm}
\includegraphics[angle=-90,width=9.0cm]{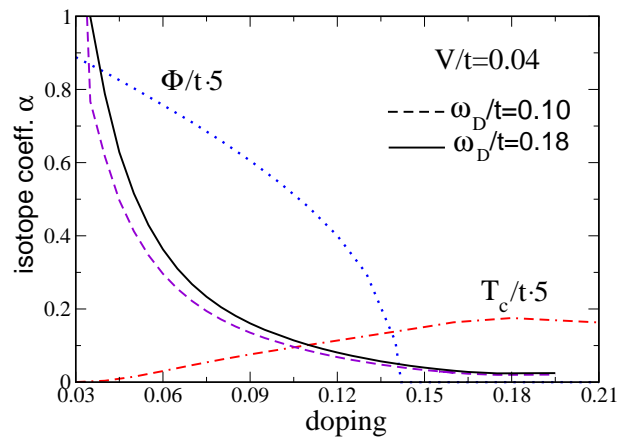}
\caption{\label{fig:3}
(Color online)
Isotope coefficient $\alpha$ as a function of doping for 
two phonon cutoffs $\omega_D$. Also shown are the curves for
$\Phi$ and $T_c$ from Fig. \ref{fig:1}}.  
\end{figure}

Fig. \ref{fig:3} shows $\alpha$ as a function of doping for $V/t=0.04$ and
two phonon cutoffs $\omega_D$ corresponding to the buckling and 
half-breathing phonon modes in YBa$_2$Cu$_3$O$_7$. \cite{Pintschovius} 
In the overdoped region
$\alpha$ is nearly independent of $\omega_D$ and $\delta$ and about
0.03, i.e., very small. In the underdoped region 
$\alpha$ monotonically increases with decreasing $\delta$
and reaches appreciable values, for instance, 1/4 at a $T_c$
which is only reduced by a factor 2 from its maximum value.

To understand the increase of $\alpha$ at low dopings better one can rewrite
Eq. (\ref{alpha}) approximately as $\langle N_d(\omega)\rangle/N_d(0)(-F_{22})$,
using a low $T_c$ approximation in the denominator.
$\langle N_d(\omega)\rangle$ is an average of $N_d(\omega)$ around the phonon
frequency $\omega_D$ over an energy
interval of about $\omega_D$. This interval is determined by the
functions $\psi$ in Eq. (\ref{F12}) and caused by the sharp cutoff in
Matsubara frequencies.
According to Fig. \ref{fig:2} $N_d(0)$ decreases rapidly with 
decreasing $\delta$
reflecting the fact that $N_d(0)$ is due to the arcs left over
from the Fermi line after formation of the $d$-CDW gap. The length of the 
arcs, however, decreases strongly with decreasing $\delta$. 
From Fig. \ref{fig:2} it is clear that
for most phonon frequencies the large spectral weight near the 
$d$-CDW gap will substantially contribute to this average. As a result 
$\langle N_d(\omega) \rangle/N_d(0) \gg 1$ and, since $-F_{22}$ is
a slowly increasing function with decreasing $\delta$, a large 
enhancement of $\alpha$ results at low dopings.  
A sharp cutoff for real frequencies, usually used in BCS-theory, 
would yield
$\langle N_d(\omega)\rangle =N_d(\omega_D)$. 
Consequently $\alpha$ would exhibit 
strong resonances for $\omega_D \sim \Phi$ or, more generally,
if $\omega_D$ is near well-pronounced peaks in $N_d(\omega)$.
Implementing the phonon cutoff in terms of Matsubara frequencies, 
as we did, corresponds to a rather soft cutoff in real frequencies.
Such a procedure is closer to an exact solution of Eliashberg equations,
yields a finite phonon contribution to $Z$ and is also free of the above
unphysical resonances in $\alpha$ if $\omega_D$ and 
the gap energy $\Phi$ are of similar magnitude. Another advantage of our 
procedure is that $\alpha$ depends only weakly on $\omega_D$, also for
$\delta \leq \delta_c$. If there is no pseudogap $N_d(\omega)$ is
rather constant in the phonon energy region which means that 
the above density ratio $\langle N_d(\omega)\rangle/N_d(0)$ is one and $\alpha$
very small for all dopings.  

Qualitatively, the curves for $\alpha$ in Fig. \ref{fig:3} are similar to 
those in Refs.
\cite{Williams,Dahm} where phenomenological
pseudogaps were used and the problem of resonances was avoided
either by considering only the limit $\omega_D \gg \Phi$ or
by a non-states-conserving pseudogap. This as well as the above
approximate expression for $\alpha$ in terms of 
the density ratio $\langle N_d(\omega)\rangle/N_d(0)$ suggests
that the above curves for $\alpha$ are rather independent of the
specific features of our model ($d$-CDW with long-range order) but rather 
generic for underdoped cuprates with a pseudogap.

Very remarkable in Fig. \ref{fig:3} is the fact that the tiny
value of 0.04 for the effective EP coupling $V/t$ is able to produce 
large values for $\alpha$ comparable to those seen in experiment in the 
underdoped region. The LDA yields $\lambda_d \sim 0.022$ 
in YBa$_2$Cu$_3$O$_7$, \cite{Heid} which 
is roughly one order of magnitude smaller than 
$\lambda_s$. \cite{Heid1,Cohen} (For a different view on the magnitude
of the EP coupling constants in cuprates, see Ref. \cite{Reznik}).
Using the LDA value 
$N_d(0) = 1.108/eV$ from Ref. \cite{Heid}, the relation Eq.(\ref{lambdad})
and $t=0.5 eV$ 
we get in the LDA $V/t \sim 0.04$ which is the value used in our calculation. 
This shows that the
large experimental values for $\alpha$ in the underdoped region
do not contradict, at least in our competing model, the
small LDA values for the EP coupling. The case of overdoped samples
is presently less clear because of conflicting experimental 
results. \cite{Franck,Pringle}
An isotope coefficient $\alpha$ which is small throughout the overdoped 
region \cite{Pringle} would
agree with our Fig. \ref{fig:3}. 

\section{Extension to finite correlation lengths of the $d$-CDW}

In the previous sections our employed mean field treatment yielded a 
$d$-CDW with 
long-range order and excitations with infinite long lifetimes. Such 
idealizations are 
certainly not realized in the cuprates and one may wonder to what 
degree our previous
results depend on them. Generally speaking we do not expect drastic 
modifications because
the strong increase of $\alpha$ with decreasing doping was due to the 
rearrangement of spectral
weight due to the $d$-CDW gap. This shift of spectral weight should not be 
seriously affected 
by fluctuations or  the loss of long-range order as long as the 
correlation length is 
sufficiently large.  In this section we will investigate this expectation on a 
more quantitative
level. To this end we will employ a model\cite{Lee1,Bartosch,Kuchinskii} 
which allows to
treat exactly a certain class of semiclassical fluctuations of the off-diagonal 
order parameter.

First we consider only the $d$-CDW part of $g^{-1}$, i.e., the first and third
rows and columns of Eq.(\ref{matrix}) where we may put $Z=1$
considering again the weak-coupling case. Transforming the part induced by the
variation of the order parameter into $\bf r$-space we get,
\begin{eqnarray}  
g_{CDW}^{-1}({\bf k}-i\nabla,i\omega_n) = \nonumber \\
\left( 
\begin{array}{c c }  
i\omega_n -\epsilon({\bf k}-i\nabla)  & -i\gamma({\bf k})\Phi_0 e^{i\beta}  \\ 
i\gamma({\bf k})\Phi_0 e^{-i\beta}   & i\omega_n -\epsilon({\bf \bar{k}}-i\nabla)
\end{array} \right),
\label{matrix1}
\end{eqnarray}
with $\beta = {\bf p}{\bf x} +\phi$ and ${\bf {\bar{k}}} = {\bf k}-{\bf Q}$. 
${\bf p}$ is a random variable with cartesian components 
distributed according to a Lorentzian. $\phi$ is a random phase which will not enter
our final expressions and thus its distribution function has not to be specified.
We apply now the unitary transformation 
\begin{equation}
 U_1({\bf x}) = e^{i\beta \sigma_3 /2},
\label{U}
\end{equation}
to Eq.(\ref{matrix1}), where $\sigma_3$ is a Pauli matrix. After some algebra we find,
\begin{eqnarray}
\tilde{g}_{CDW}^{-1}\hspace{-0.3cm}&(&\hspace{-0.3cm}{\bf k}-i\nabla,i\omega_n) = 
U^\dagger_1({\bf x}) g_{CDW}^{-1}({\bf k}-i\nabla,i\omega_n) U_1({\bf x}) \nonumber \\
&=&\left( 
\begin{array}{c c }  
i\omega_n -\epsilon({\bf k}) +\epsilon_1  & -i\gamma({\bf k})\Phi_0   \\ 
i\gamma({\bf k})\Phi_0    & i\omega_n -\epsilon({\bf \bar{k}}) +\epsilon_2  
\end{array} 
\right),\nonumber \\
\label{matrix2}
\end{eqnarray}
with the abbreviations $\epsilon_1 = {\bf v}({\bf k}){\bf p}/2 $ and 
$\epsilon_2 = -{\bf v}({\bf {\bar{k}}}){\bf p}/2$. $v_{x,y}({\bf k})$ are given by 
$\partial \epsilon({\bf k})/\partial {k_{x,y}}$. In deriving
Eq.(\ref{matrix2}) we also expanded  $\epsilon({\bf k}-i\nabla)$ up to first order
in $\nabla$ assuming that the changes in one-particle energies induced by the
variation of the order parameter vary slowly in space.
To simplify the following we will take
$\epsilon_1=\epsilon_2=\epsilon$ which holds exactly for the case $t'=0$.

The matrix in Eq.(\ref{matrix2}) can be diagonalized by a second unitary 
transformation $U_2$
yielding,
\begin{eqnarray}
 U_2^\dagger({\bf k}) \tilde{g}_{CDW}^{-1}({\bf k},i\omega_n) U_2({\bf k}) = \nonumber \\
\left(
\begin{array}{c c }  
i\omega_n -\lambda_1({\bf k}) +\epsilon  &     0   \\ 
        0                            & i\omega_n -\lambda_2({\bf k}) +\epsilon  
\end{array} 
\right),
\label{U2}
\end{eqnarray}
with the eigenvalues $\lambda_{1,2}({\bf k})$ of the unperturbed $d$-CDW, 
given by
Eq.(\ref{Pole1}).
Taking also superconductivity into account we note that the Heisenberg
interaction is invariant under the gauge transformation $U_1$. Applying the
second unitary transformation $U_2$ to the Heisenberg interaction
and using the BCS factorization
the transformed matrix $g^{-1}$ splits into two 2x2 matrices and the 
resulting gap equation can easily be calculated. Adding also phonons one
finds that Eqs.(\ref{Tc})-(\ref{F12}) still hold if the density 
$N_d(\omega)$ of 
Eq.(\ref{N}) is replaced by the expression,
\begin{equation}
\frac{\omega}{\pi N_c} \sum_{{\bf k},\beta} \gamma({\bf k})^2
\Im \frac{1}{(\omega-i\eta+\epsilon)^2-\lambda_\beta^2({\bf k})}.
\label{N2}
\end{equation}
Finally,
\begin{figure} 
\vspace*{-0.2cm}
\includegraphics[angle=270,width=9.0cm]{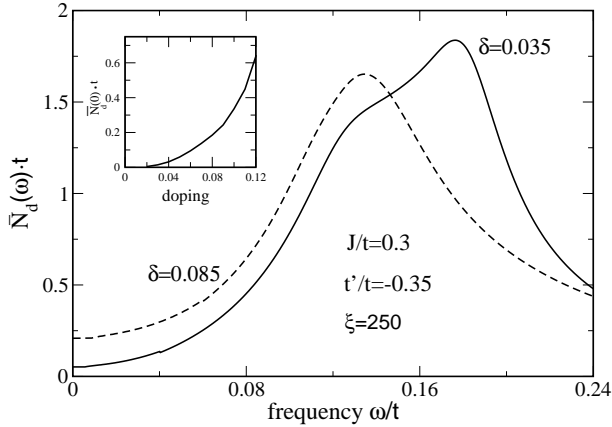}
\caption{\label{fig:4}
Weighted density of electronic states $\bar{N}_d(\omega)$ for two 
dopings $\delta$ for a correlation length $\xi = 250$. 
The inset shows $\bar{N}_d(0)$ at low dopings.
}
\end{figure}
Eqs.(\ref{Tc})-(\ref{F12}) have to be averaged over the distribution 
function $P({\bf p})$,
\begin{equation} 
P({\bf p}) = \frac{\kappa^2}{\pi^2(p_x^2 + \kappa^2)(p_y^2+\kappa^2)},
\end{equation}
where $\kappa = 1/\xi$ and $\xi$ is the correlation length of the off-diagonal
order parameter fluctuations. According to Eq.(\ref{matrix1}) $\Phi_0$
is the equilibrium $d$-CDW order parameter after omitting the factor 
$\gamma({\bf k})$. It thus varies in $\bf r$-space with the momentum $\bf Q$
because it connects electron states with momenta $\bf k$ and $\bar{\bf k}$.
The random variable $\bf p$ modulates the equilibrium total momentum $\bf Q$
of the $d$-CDW in an additive way and it is distributed according to a 
Lorentzian. It is sufficient to apply the necessary average over  
$\bf p$ just to the expression of Eq.(\ref{N2}) yielding,
\begin{eqnarray} 
\bar{N}_d(\omega) = \frac{\omega}{\pi N_c} \sum_{{\bf k},\beta} \gamma({\bf k})^2
\nonumber \\
\Im \frac{1}{(\omega-i/(2\xi)(|v_x({\bf k})|+|v_y({\bf k})|))^2 -\lambda_\beta^2({\bf k})}.
\label{N1}
\end{eqnarray}
It is easy to see that the above
expression can formally be obtained from $N_d(\omega)$ if one 
replaces there
the infinitesimal $\eta$ by the finite, $\bf k$-dependent imaginary part 
$ (|v_x({\bf k})|+|v_y({\bf k})|)/(2\xi)$. The latter quantity is in 
Eq.(\ref{N1}) averaged over
the Fermi line in the pure $d$-CDW state, i.e., for $\beta =1$ over the arc 
around the nodal line
and for $\beta = 2$ over the remaining small piece near the antinodal point 
which however, vanishes
for our two considered dopings. Thus the $\beta = 2$ contribution in Eq.(\ref{N1}) is 
negligible small. In the $\beta = 1$ contribution 
$ (|v_x({\bf k})|+|v_y({\bf k})|)/(2\xi)$
varies only little along the arc so we may replace this quantity by its average over the
arc and obtain for $\delta = 0.085$ the value  $1.15 v_F/\xi$ 
where $v_F \sim 4.6t$ is
the square root of the Fermi surface average of ${\bf v}^2({\bf k})$ in 
the normal state. This allows to describe phase
correlations with a finite correlation length $\xi$ as an inverse life time effect with
energy $\Gamma/t = 1.15 v_{F}/(t\xi)$.
\begin{figure} 
\vspace*{-0.2cm}
\includegraphics[angle=270,width=9.0cm]{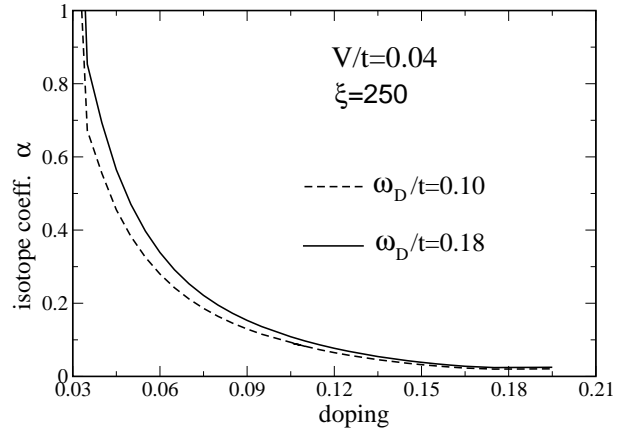}
\caption{\label{fig:5}
Isotope coefficient $\alpha$ as a function of doping for 
two phonon cutoffs $\omega_D$ calculated in the presence of 
short-ranged correlated phase fluctuations with correlation length 
$\xi = 250$. 
}
\end{figure}

Fig. 4 shows $\bar{N}_d(\omega)$ for the two dopings of Fig. 2
and for the case of a correlation length $\xi = 250$, corresponding to an 
inverse lifetime of about $0.02t$. Though such a large correlation length
may seem to simulate a rather well-ordered state most of the fine structures in
Fig. 2 are wiped out by phase fluctuations.  In particular, the two peaks 
seen in Fig. 2
have merged into one broad and rather structureless peak. At low 
frequencies the changes introduced by phase fluctuations are rather minor.  
In the inset
in Fig. 4 the static value $\bar{N}_d(0)$ is plotted as a function of doping
showing the pronounced decrease of $\bar{N}_d(0)$ with decreasing $\delta$
similar as in the case without phase fluctuations in Fig. 2. This behavior 
for the density is rather robust as function of $\xi$ as long as
$\xi \gg 1$ holds. 

Fig. 5 shows the doping dependence of the isotope coefficient $\alpha$ 
using the same parameter
as in Fig. 3 but $\bar{N}_d(\omega)$ instead of $N_d(\omega)$. 
Comparing Figs. 3 and 5 reveals
that this change of densities does hardly affects $\alpha$ so that 
corresponding curves in these
two figures are practically identical. This demonstrates that
the steep increase of $\alpha$ with decreasing doping is not related or 
even caused by
the sharp peaks present in $N_d(\omega)$ or by special values for the 
phonon cutoff $\omega_D$.
Instead, it is a rather universal property caused by the large shift of 
spectral weight towards higher frequency due to the pseudogap. 
If the correlation length $\xi$ is decreased from large to small values of the
order of the lattice constant 
the depletion in the density $\bar{N}_d(0)$ at 
low energies becomes smaller and smaller. In accordance with the decreasing
shift of spectral weight from low to high energies the isotope coefficient 
$\alpha$ also decreases approaching a similar small value as
in the absence of a pseudogap, i.e., in the overdoped region.

Fig. 5 can qualitatively be understood in a simple way considering our 
previously approximate expression $\langle N_d(\omega)\rangle/{N_d(0)}$ 
for $\alpha$. The numerator 
is essentially given by the area under the density curve in Figs. 2 and 4. 
Its value thus
is independent of the shape of the  density curve, i.e., whether it has sharp or
broad  peaks, as long as the area below the curve is constant. This area 
is practically
the same in Figs. 2 and 4.  On the other hand the $\omega =0$ values of 
the densities 
both decrease strongly
and in a similar way with decreasing $\delta$.  As a result $\alpha$ 
should be of similar
magnitude in both cases and, in particular, show a strong increase 
towards low dopings 
in agreement with Fig. 5. The above approximate expression for 
$\alpha$ may also explain 
why the calculated values for $\alpha$ of Ref. \cite{Jarrell} are for our 
electron-phonon
coupling $V$ much smaller than ours: Their density of 
state function at $\lambda_d = 0$, 
plotted in their Fig. 4, is large at $\omega=0$ compared to the modulation 
due to
the pseudogap which implies only a small redistribution of spectral weight
by the pseudogap. 

\section{Conclusions}

We have shown that a mean-field treatment of the $t$-$J$ model which
identifies the pseudogap with the gap of a $d$-CDW state is able
to explain the large isotope effect in underdoped cuprates.
Interestingly, 
a very small EP coupling constant $V/t \sim 0.04$ is sufficient
to explain the experimental data. This value is very close to that calculated
for YBa$_2$Cu$_3$O$_7$ within the LDA. This shows that the large values for
$\alpha$ of order 1 found in the underdoped region are, at least in our
competing model, compatible with the small EP coupling constants predicted
by the LDA. 
The obtained huge increase of the isotope coefficient $\alpha$ with
decreasing doping is rather independent of the phonon cutoff frequency
$\omega_D$ and the spectral properties of the excitations in the pseudogap
state. The latter information is obtained by considering a simple phase 
fluctuation model where the $d$-CDW state has only short-range correlations. 

Most important for the large increase of $\alpha$ with decreasing doping 
is in our calculation the large depletion of spectral weight at low frequencies
and its shift to high energies by the pseudogap. Because of this we conjecture
that our results are not specific to the employed $d$-CDW 
providing the pseudogap but also
hold for other order parameters such as the antiferromagnet order parameter
as long as they lead to a strong depletion of spectral weight at low 
frequencies.  
Our calculation shows, in particular, that it is not necessary
to assume a large EP interaction in cuprates or extrinsic effects such as
pairbreaking due to impurities \cite{Tallon,Bill} 
to explain the observed isotope effect of $T_c$. 

The authors thank O. Gunnarsson for discussions and A.M. Ole\'s for a
critical reading of the manuscript.


\begin{thebibliography}{99}
\bibitem{comment}
      In most $1/N$ treatments, for instance in Refs. \cite{Grilli,Greco},
      the prefactor 2 in the scaling of coupling constants is omitted.
      The physical Hamiltonian is thus for $N=2$ reproduced only with
      an overall prefactor 1/2 causing additional prefactors of 1/2
      in front of every coupling constant in subsequent equations.
\end{thebibliography}
\end{document}